\documentclass[prb,twocolumn,showpacs,superscriptaddress,preprintnumbers]{revtex4}
\usepackage{graphicx}
\usepackage{dcolumn}
\usepackage{bm}

\begin{document}

\title{Electrodynamics of the Nodal Metal in Weakly Doped High-$T_{c}$
Cuprates}
\author{Y. S. Lee}
\thanks{Present address: Spin Superstructure Project, ERATO, JST, c/o AIST
Tsukuba central 4, 1-1-1 Higashi, Tsukuba 305-8562, Japan}
\affiliation{Department of Physics, University of California at San Diego, La Jolla,
California 92093-0319, USA}
\author{Kouji Segawa}
\affiliation{Central Research Institute of Electric Power Industry, Komae, Tokyo
201-8511, Japan}
\author{Z. Q. Li}
\affiliation{Department of Physics, University of California at San Diego, La Jolla,
California 92093-0319, USA}
\author{W. J. Padilla}
\thanks{Present address: Los Alamos National Laboratory, MSK764, MST-10, Los
Alamos, NM 87545, USA.}
\affiliation{Department of Physics, University of California at San Diego, La Jolla,
California 92093-0319, USA}
\author{M. Dumm}
\thanks{Present address: Physikalisches Institut, Universit\"{a}t Stuttgart,
70550 Stuttgart, Germany}
\affiliation{Department of Physics, University of California at San Diego, La Jolla,
California 92093-0319, USA}
\author{S. V. Dordevic}
\affiliation{Department of Physics, Brookhaven National Laboratory, Upton, New York
11973, USA}
\author{C. C. Homes}
\affiliation{Department of Physics, Brookhaven National Laboratory, Upton, New York
11973, USA}
\author{Yoichi Ando}
\affiliation{Central Research Institute of Electric Power Industry, Komae, Tokyo
201-8511, Japan}
\author{D. N. Basov}
\affiliation{Department of Physics, University of California at San Diego, La Jolla,
California 92093-0319, USA}
\date{\today }

\begin{abstract}
We report on the detailed analysis of the infrared (IR) conductivity of two
prototypical high-$T_{c}$ systems YBa$_{2}$Cu$_{3}$O$_{y}$ and La$_{2-x}$Sr$%
_{x}$CuO$_{4}$ throughout the complex phase diagram of these compounds. Our
focus in this work is to thoroughly document the electromagnetic response of
the nodal metal state which is initiated with only few holes doped in parent
antiferromagnetic systems and extends up to the pseudogap boundary in the
phase diagram. The key signature of the nodal metal is the two-component
conductivity: the Drude mode at low energies followed by a resonance in
mid-IR. The Drude component can be attributed to the response of coherent
quasiparticles residing on the Fermi arcs detected in photoemission
experiments. The microscopic origin of the mid-IR band is yet to be
understood. A combination of transport and IR data uncovers fingerprints of
the Fermi liquid behavior in the response of the nodal metal. The
comprehensive nature of the data sets presented in this work allows us to
critically re-evaluate common approaches to the interpretation of the
optical data. Specifically we re-examine the role of magnetic excitations in
generating electronic self energy effects through the analysis of the IR
data in high magnetic field.
\end{abstract}

\pacs{74,25.Gz, 74,72.Bk}
\maketitle

\section{Introduction}

Recent experimental studies of weakly doped cuprates have revealed a novel
enigmatic state: the nodal metal.\cite%
{Ando01,Zhou03,Dumm03,Yslee04,taillefer03} This state is realized with only
few holes introduced in parent antiferromagnetic (AF) insulators where doped
materials still show AF ordering or the so-called spin-glass regime at low
temperatures and extends throughout the notorious pseudogap region.
Spectroscopic signatures of this electronic state include a Drude-like
response in the optical conductivity\cite{Dumm03,Yslee04} and a
quasiparticle (QP) peak on the nodal \textquotedblleft Fermi
arc\textquotedblright\ seen in the photoemission data.\cite{Yoshida03} The
electronic mobility and the Fermi velocity of nodal QP, as well as their
optical effective mass, remain virtually unchanged as doping progresses to
much higher carrier density where superconductivity is optimized.\cite%
{Ando01,Zhou03,Padilla04} All of these observations contradict a common
characterization of the AF-ordered region of phase diagram as
\textquotedblleft antiferromagnetic insulator\textquotedblright .

Carrier dynamics in the nodal metal is strongly influenced by unidirectional
spin and/or charge self-organization effects commonly referred to as stripes.%
\cite{ichikawa00} Unambiguous manifestations of stripes can be found through
experiments probing the anisotropy of conductivity within the nearly square
CuO$_{2}$ planes.\cite{ando02} Infrared (IR) experiments carried out with
polarized light reveal the enhancement of optical conductivity along the
\textquotedblleft spin stripes\textquotedblright\ direction in La$_{2-x}$Sr$%
_{x}$CuO$_{4}$ (LSCO)\cite{ando02,Dumm03} and along the ripples of the
electron density in the CuO$_{2}$ planes in YBa$_{2}$Cu$_{3}$O$_{y}$ (YBCO).%
\cite{Yslee04}

The carrier density $n$ in the nodal metal varies linearly with doping: a
result that became established shortly after the discovery of high-$T_{c}$
superconductors\cite{tagaki89} and detailed in more recent studies.\cite%
{Ando01,segawa04,ando04} An important consequence of the $n\propto x$
relationship is a dramatic suppression of the resistivity probed with
currents along the CuO$_{2}$ plane as one proceeds from
antiferromagnetically ordered phases to over-doped metals. It is rather
surprising that dynamical characteristics of charges including the electric
mobility,\cite{Ando01} the nodal Fermi velocity,\cite{Zhou03} or
quasiparticles optical mass $m^{\ast }$\cite{Padilla04} reveal negligible
changes throughout the phases diagram.

In this paper we report on a detailed study of the nodal metal state in
YBCO: a prototypical high-$T_{c}$ superconductor carried out using untwinned
single crystals for ten different dopings between $y$ = 6.28 and $y$ = 7.00.
These new results unveil the evolution of the electromagnetic response as
doping progresses from AF-ordered phase to superconducting materials. New
data allow us to comprehensively test and re-examine common approaches to
analyze the electrodynamics of cuprates. We show that the IR signatures of
the nodal metal involve a two-component conductivity: the Drude mode at low
energies followed by a resonance in mid-IR. The Drude component can be
attributed to the response of coherent QP residing on the Fermi arcs; the
microscopic origin of the mid-IR band is yet to be thoroughly understood.
The two-component conductivity extends to the pseudogap boundary in the
phase diagram at $T^{\ast }$. We find similar trends in the response of the
LSCO system and conclude that the nodal metal state is a hallmark of the
electrodynamics of the pseudogap regime. In order to narrow down the range
of possible microscopic interpretations of the mid-IR structure we have
carried out studies of the $y$ = 6.65 single crystal in high magnetic field (%
\textbf{H}). These experiments allow one to assess the possible role of
magnetic fluctuation in the QP dynamics. Our findings are at odds with the
idea of the dominant role of magnetic resonance seen in neutron scattering
experiments in mid-IR response.

This paper is organized as follows: In Sec. II we report on experimental
details of crystal preparation, characterization, and reflectance
measurements. The DC resistivity, the Hall coefficient, and the raw
reflectivity data for YBCO detwinned single crystals are also discussed in
this section. Section III is devoted to the survey of the optical
conductivity with variation of doping and temperature. Characteristic
features of the conductivity are discussed in the context of the phase
diagram of YBCO system. Section IV is focused on the analysis on the optical
conductivity in terms of two competitive approaches: single-component and
multi-component descriptions. A simple two-component model is shown to grasp
the gross trends in the evolution of the conductivity throughout the entire
doping and temperature range. In Sec. V we investigate the transformation of
the optical conductivity below $T_{c}$. It is found that the dominant
contribution to the superconducting condensate originates from the coherent
quasiparticles on the Fermi arc. In Sec. VI we discuss measurements in high
magnetic fields carried out for $y$ = 6.65 crystal with the goal to address
the role of the magnetic resonance in the CuO$_{2}$ plane optical
conductivity of YBCO. Changes of reflectivity spectra with magnetic field
are simulated within the framework of the electron-boson coupling theory,
and compared with experimental results. In Sec. VII we analyze some of the
implications of new experimental results focusing on common trends between
the YBCO and LSCO series as well as on the possible origin of the mid-IR
absorption. Summary and outlook of this work are presented in Sec. VIII.

\section{Experimental details and raw data}

We investigated detwinned YBCO single crystals with oxygen content $y$ =
6.28, 6.30, 6.35, 6.40, 6.43, 6.50, 6.55, 6.65, 6.75, and 7.00. A summary of
the N\'{e}el temperatures $T_{N}$ for AF ordered samples and of
superconducting transition temperatures $T_{c}$ for superconducting
specimens is given in Table~I.\cite{Segawa01,lavrov99} Single crystals were
grown by a conventional flux method in Y$_{2}$O$_{3}$ crucibles and
detwinned under uniaxial pressure at CRIEPI.\cite{Segawa01} Note that the
quality of these crystals is comparable to those grown in BaZrO$_{3}$
crucibles, as is evidenced by the thermal conductivity data below $T_{c}$.%
\cite{sun04} Annealing under uniaxial pressure also aligns chains fragments
along the $b$-axis in non-superconducting YBCO ($y$ = 6.28 - 6.35). The
typical sizes of the nearly rectangular samples are $\sim 1\times 1.5$ mm$%
^{2}$ in the $ab$ plane with the sub-millimeter thickness. Detwinned single
crystals allow us to investigate the response of the CuO$_{2}$ plane ($a$%
-axis) unperturbed by contributions due to the chain structures along the $b$%
-axis. In this paper, we restrict our study to the $a$-axis optical spectra;
we reported on the electrodynamics associated with the chain segments in
heavily underdoped YBCO in two other publications.\cite{Yslee04,ylee05}

\begin{figure}[tbp]
\includegraphics[width=0.37\textwidth]{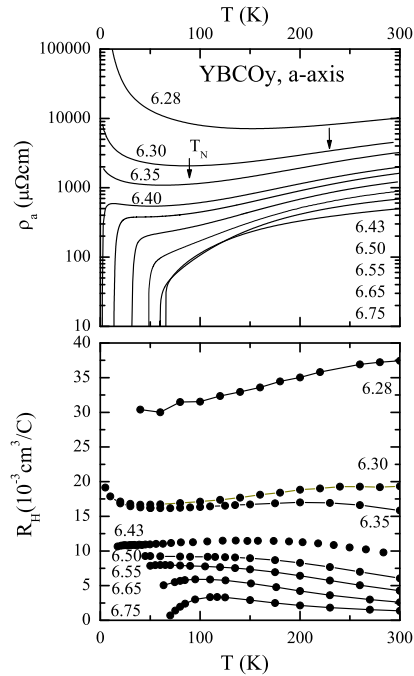}
\caption{(a) Dc resistivity and (b) Hall coefficient curves for $y$ = 6.28 -
6.75 single crystals. The arrows represent $T_{N}$ for AF samples. The $%
T_{N} $ for $y$ = 6.28 is above 300 K.}
\end{figure}

\begin{table}[tbp]
\caption{Summary on transition temperatures of the YBCO samples analyzed in
this paper. N\'{e}el temperature $T_{N}$ is determined from the c-axis
resistivity measurement [A.N. Lavrov \textit{et al}., Phys. Rev. Lett. 
\textbf{83}, 1419 (1999)]. Superconducting transition temperature $T_{c}$ is
determined by the onset temperature of $\protect\rho _{\text{dc}}(T)=0$.}%
\begin{ruledtabular}
\begin{tabular}[t]{ccccccccccc}
$y$ & 6.28 & 6.30 & 6.35 & 6.40 & 6.43 & 6.50 & 6.55 & 6.65 & 6.75 & 7.00 \\
\hline
$T_{N}$ (K) & >300 & 230 & 80 & & & & & \\
\hline
$T_{c}$ (K) & & & & 2 & 13 & 31 & 48 & 60 & 65 & 92%
\end{tabular}
\end{ruledtabular}
\end{table}

The DC resistivity $\rho _{\text{dc}}(T)$ and Hall coefficient $R_{H}(T)$
curves of the YBCO samples studied in this paper are displayed in Fig. 1.%
\cite{Segawa01,segawa04} At moderately high $T$ $\rho _{\text{dc}}$\ shows a
metallic behavior in all samples including those that order
antiferromagnetically ( $y$ < 6.40) [top panel].\cite{Ando01}
This metallic behavior persists well below $T_{N}$. At low $T$ one can
notice a negative slope of the resistivity in both non-superconducting
samples ($y=6.28$ - $6.35$) as well as in superconducting specimens ($y=6.40$
and $6.43$). Resistivities for $y$ = 6.65 and 6.75 samples overlap between
70 and 140 K; this is a peculiar behavior associated with the so-called 60
K-phase anomaly.\cite{Segawa01} In the underdoping range the Hall
coefficient does not show a strong $T$ dependence and therefore data in the
bottom panel of Fig.~1 can be used for obtaining a rough estimate of the
carrier density from $R_{H}(T)$ ($R_{H}$ = 1/$ne$).

\begin{figure}[tbp]
\includegraphics[width=0.46\textwidth]{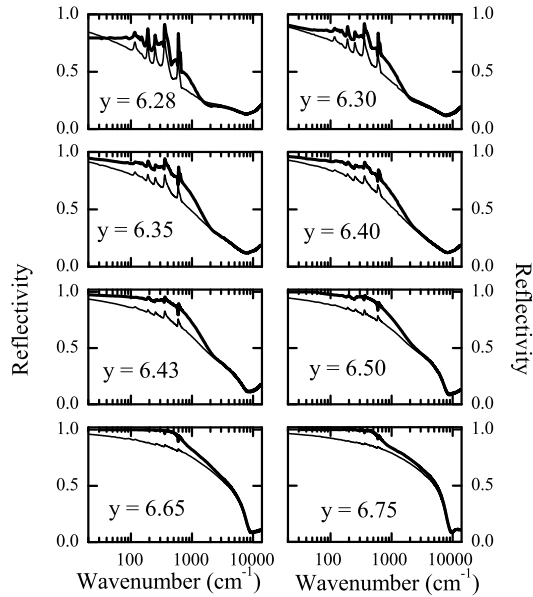}
\caption{Reflectance spectra for $y$ = 6.28 - 6.75 single crystals measured
at 10 K (thick lines) and 293 K (thin lines). For clarity, reflectance
spectra are shown in the range from 0 to 1.02 on the vertical axis for
highly reflective samples with $y$ $\geq 6.43$.}
\end{figure}

Reflectivity spectra $R(\omega )$ at nearly normal incidence were measured
with polarized light at frequencies from 20 to 48000 cm$^{-1}$ and at
temperatures from 10 to 293 K at UCSD. A \textit{in-situ} gold-coating
technique was employed for reference spectra.\cite{homes93-1} Far-IR
reflectance of actual gold films $R^{gold}(\omega )$ was evaluated from
measurements of DC resistivity using the Hagen-Rubens formula and mid- and
near-IR reflectivity of the gold coatings was directly determined using
ellipsometric measurements. $T$-dependence of $R^{gold}(\omega )$ has to be
taken into account for calibration in far-IR range. Thus the obtained data
for $R^{gold}(\omega )$ at various temperatures were used to generate
absolute reflectivity for all studied samples.

In Fig. 2 we show $R(\omega )$ for polarization of the \textbf{E} vector
along the $a$-axis measured at 10 K and 293 K. A common feature of $R(\omega
)$ traces is their `metallic character': all spectra reveal an increase of
the absolute value with lowering $\omega $ starting from a `plasma minimum'
at $\simeq $ 10,000 cm$^{-1}$. The magnitude of the reflectance in far-IR
increases with doping so that the contribution of the transverse optical
phonon modes becomes less pronounced. With $T$ decreasing, the low frequency 
$R(\omega )$ increases significantly, whereas the change in the mid-IR is
fairly small. For $y$ = 6.28 and 6.30 where the low $T$ resistivity shows a
semiconducting behavior below $T_{Loc}$, spectra measured at 10 K and 80 K
cross at the lowest frequencies (80 K data are not shown). As will be
detailed in Sec. III, this latter effect is due to the modification of
charge dynamics by disorder.

\begin{figure}[tbp]
\includegraphics[width=0.48\textwidth]{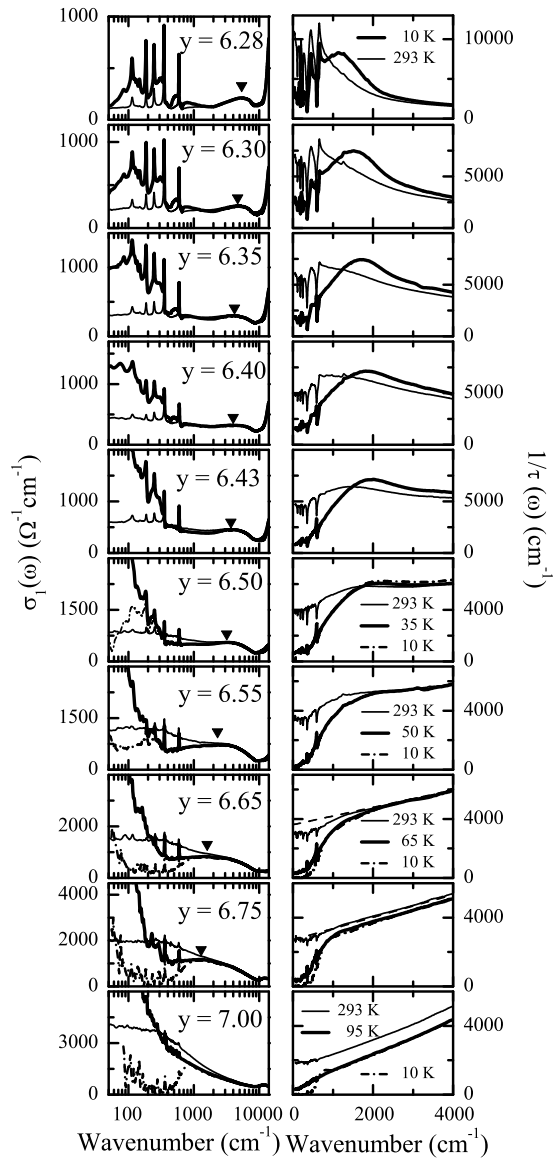}
\caption{Spectra of the $a$-axis conductivity $\protect\sigma _{1}(\protect%
\omega )$ [left panels] and of the scattering rate 1/$\protect\tau (\protect%
\omega )$ [right panels] extracted from the extended Drude model as
described in the text for a series of YBCO single crystals. Thick solid
lines show the data at lowest temperature in the normal state: 10 K for
non-superconducting crystals $y$ = 6.28 - 6.40 and $T$ $\sim T_{c}$ for
superconducting compounds $y$ = 6.43 - 7.00. The solid triangles in the left
panels mark the positions of the mid-IR absorption bands. In right panles
for $y$ = 6.65 and 6.75, the dashed lines represent the linear $\protect%
\omega $-dependence.}
\end{figure}

\section{Doping trends in the optical conductivity}

The complex optical conductivity spectra $\widetilde{\sigma }(\omega
)=\sigma _{1}(\omega )+i\sigma _{2}(\omega )$ were obtained from $R(\omega )$
using the Kramers-Kronig (KK) transformation. The KK-derived results are
consistent with those determined independently by ellipsometry in the
near-IR and visible regions. These results are presented in the left panels
of Fig. 3 where we plot the $a$-axis $\sigma _{1}(\omega )$ spectra at 10 K
and 293 K. For superconducting crystals we also show data at $T\sim T_{c}$.
The sharp peaks below 700 cm$^{-1}$ are due to transverse optical phonon
modes. In all samples we find evidence for a charge transfer (CT) gap near
10,000 cm$^{-1}$; this feature is especially evident in most weakly doped
compounds, but can be recognized even in the data for the $y$ = 6.75
crystal. The intra-gap conductivity of all materials also reveals common
patterns that become particularly clear in the low-$T$ spectra (thick
lines). We find that the low-$T$ spectra in the heavily underdoped region
are composed of two separate absorption features: a coherent mode ( at $%
\omega <600$ cm$^{-1}$) followed by a mid-IR absorption at 0.5 - 0.6 eV.\cite%
{Yslee04} The mid-IR structure is virtually $T$-independent, whereas the
coherent mode significantly narrows at low $T$. As the doping increases the
mid-IR absorption gradually shifts to lower frequency from $\sim $ 5,000 cm$%
^{-1}$ in the $y=6.28$ sample down to $\sim $1,300 cm$^{-1}$ for $y=6.75$,
still distinguished from the well-developed coherent mode.

\begin{figure*}[tbp]
\includegraphics[width=7in, height=3.8in]{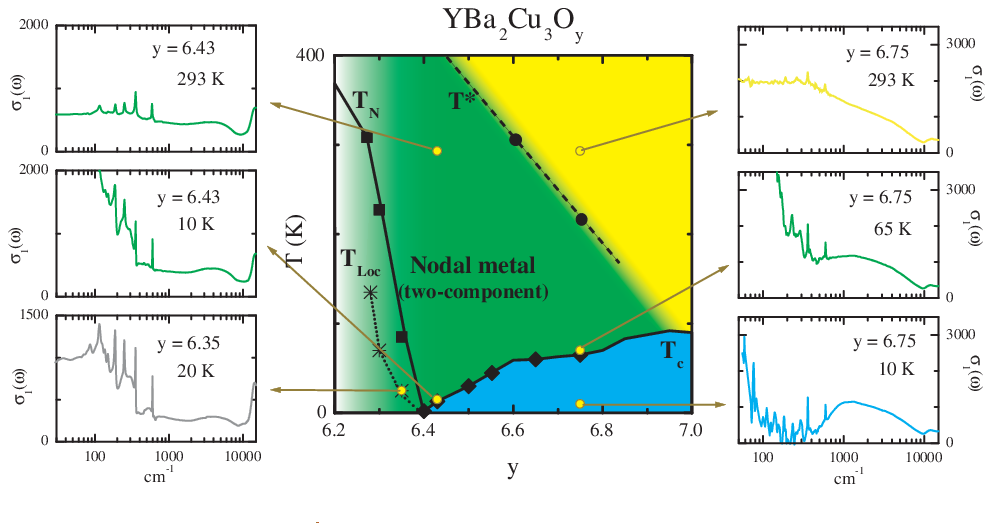}
\caption{(color online) Mapping of the representative optical spectra to the
phase diagram in a range from the heavily underdoped to the pseudogap state.
The yellow circles represent $y$ and $T$ for the corresponding spectra.
Black symbols represent transition temperatures of the samples studied in
the paper. $T_{Loc}$ is a crossover temperature signifying a semiconducting
upturn in $\protect\rho _{a}(T)$. It is noted that $T_{Loc}$ is much below $%
T_{N}$, which indicates that the semiconducting behavior is irrelevant to a
long range of the AF-ordered state.}
\end{figure*}

A remarkable result presented in Figs.~3 and 4 is a conventional Drude
behavior of an antiferromagnetically ordered crystal with $y$ = 6.35.\cite%
{Yslee04} (The details of the Drude fits are presented in Fig.~6.) This
result conflicts with a common reference to weakly doped phases as
`antiferromagnetic \textit{insulators}' since the conventional Drude
response is a standard characteristic of metallic transport. The weak upturn
in $\rho _{\text{dc}}(T)$ at $T<20$ K [Fig. 1] is most likely caused by
disorder-induced localization. Signatures of localization can be identified
in the $\omega $-dependence of the conductivity measured for samples with $%
y\leq 6.30$, where the coherent contribution to the conductivity reveals a
peak at finite far-IR frequencies.\cite{Dumm03} Importantly, the response of
all crystals remains gapless down to the lowest $\omega $ as evidenced by a
substantial spectral weight in the $\sigma (\omega )$ down to $\omega
\rightarrow 0$. It is also found that at the lowest frequencies ($\sim $ 40
- 50 cm$^{-1}$) the optical conductivity is in good agreement with the $\rho
_{\text{dc}}$\ values. This agreement holds at all temperatures. Based on
this agreement with the transport data we conclude that no dramatic changes
are likely to occur in the IR data at frequencies below our lower cut-off.
For this reason we believe that the negative slope of the low $T$
resistivity in weakly doped YBCO is not due to the opening of the insulating
gap. Instead we attribute this behavior to a band-like response even in
lightly doped samples that is modified by weak localization.\cite%
{basov94,basov98,dumm02,alternative-peak}

\begin{figure}[tbp]
\includegraphics[width=0.42\textwidth]{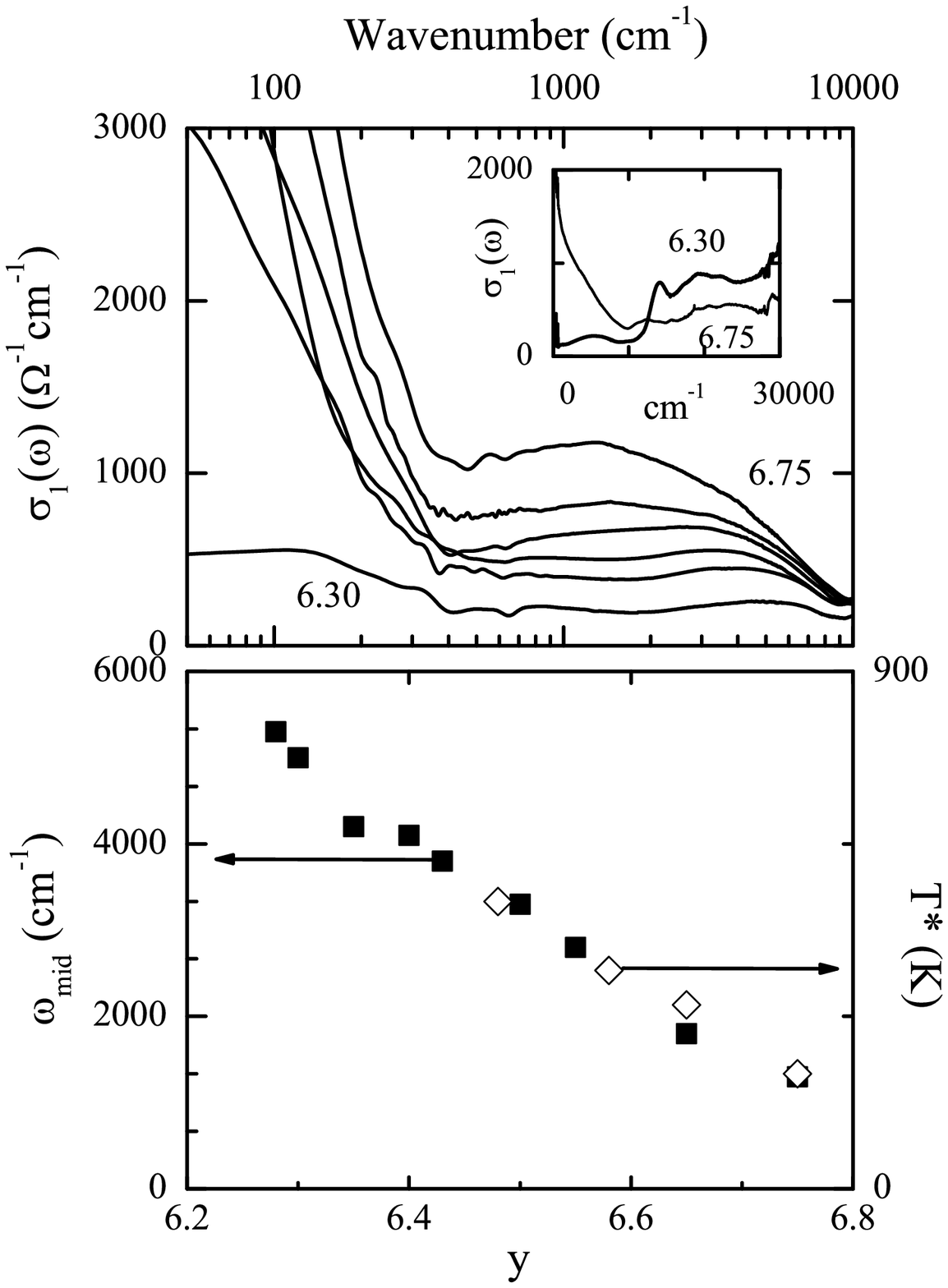}
\caption{(Top panel) Doping dependent $\protect\sigma _{1}(\protect\omega )$
at 10 K or at $T$ $\simeq T_{c})$ for $y=6.30$, 6.35, 6.43, 6.50, 6.55, and
6.75. For clarity, the sharp phonon structures are removed. Inset shows the $%
\protect\sigma _{1}(\protect\omega )$ up to 30,000 cm$^{-1}$ for $y$ = 6.30
and 6.75. (Bottom panel) Doping dependences of the peak position of mid-IR
absorption $\protect\omega _{\text{mid}}$ (solid square) and the pseudogap
onset temperature $T^{\ast }$ (open square) quoted from Ref. 22. The $%
\protect\omega _{\text{mid}}/k_{B}T^{\ast }$ values are estimated to be 7 -
9.}
\end{figure}

It is instructive to discuss the evolution of the conductivity spectra both
with temperature and doping in the context of the phase diagram of YBCO
system. The pseudogap boundary at $T^{\ast }$ is associated with the
formation of the partial (incomplete) gap in the spectrum of the low-energy
excitations.\cite{timusk99} Spectroscopic signatures of the pseudogap are
most clearly seen in the interplane $c$-axis optical conductivity showing a
depletion of the far-IR spectral weight below $T^{\ast }$.\cite%
{homes93,basov94prb} Changes of the in-plane response probed in the
polarization \textbf{E} $\parallel $ CuO$_{2}$ attributable to the pseudogap
are much more subtle. These changes are usually discussed using the extended
Drude formalism that we will introduce in Section IV. However both the
comprehensive nature of the data set presented in Fig. 3 and the quality of
these spectra allow one to identify important trends \textit{directly} in $%
\sigma _{1}(\omega )$ spectra. One difference from the earlier studies\cite%
{basov96,Puchkov96} is that we are now confident to display $T$-dependent
spectra up to 2 eV owing to both improvements in reflectance measurements
and to availability of ellipsometric results in near-IR and visible ranges.
At $T>T^{\ast }$ we observe a broad spectrum extending from $\omega =0$ to
the CT frequency (top right panel of Fig. 4). This response can be
characterized as single-component since multiple absorption features cannot
be unambiguously singled out. Crossing the $T^{\ast }$ boundary in Fig.~4
either vertically (by changing temperature at constant $y$) or horizontally
(by changing doping at constant $T$) yields similar results. Indeed, two
distinct absorption structures are affiliated with the entire green (or
gray) region of the phase diagram. These structures include a Drude mode at
the lowest frequencies and mid-IR band. Similarities are especially clear
when the low-$T$ data are compared (the middle panels of Fig .4). The main
distinction in the latter spectra taken for $y$ = 6.43 and 6.75 compounds is
in the frequency position of the mid-IR band at $\omega _{\text{mid}}$. The
softening of this feature with increasing doping is continuous as clearly
seen in Figs. 3 and 5. As pointed out above, this two-component response is
among the signatures of the nodal metal detected in AF sector of the phase
diagram (bottom left panel in Fig.~4). We therefore conclude that the nodal
metal characteristics extend throughout both the AF and pseudogap regions of
the phase diagram.

\begin{figure}[tbp]
\includegraphics[width=0.48\textwidth]{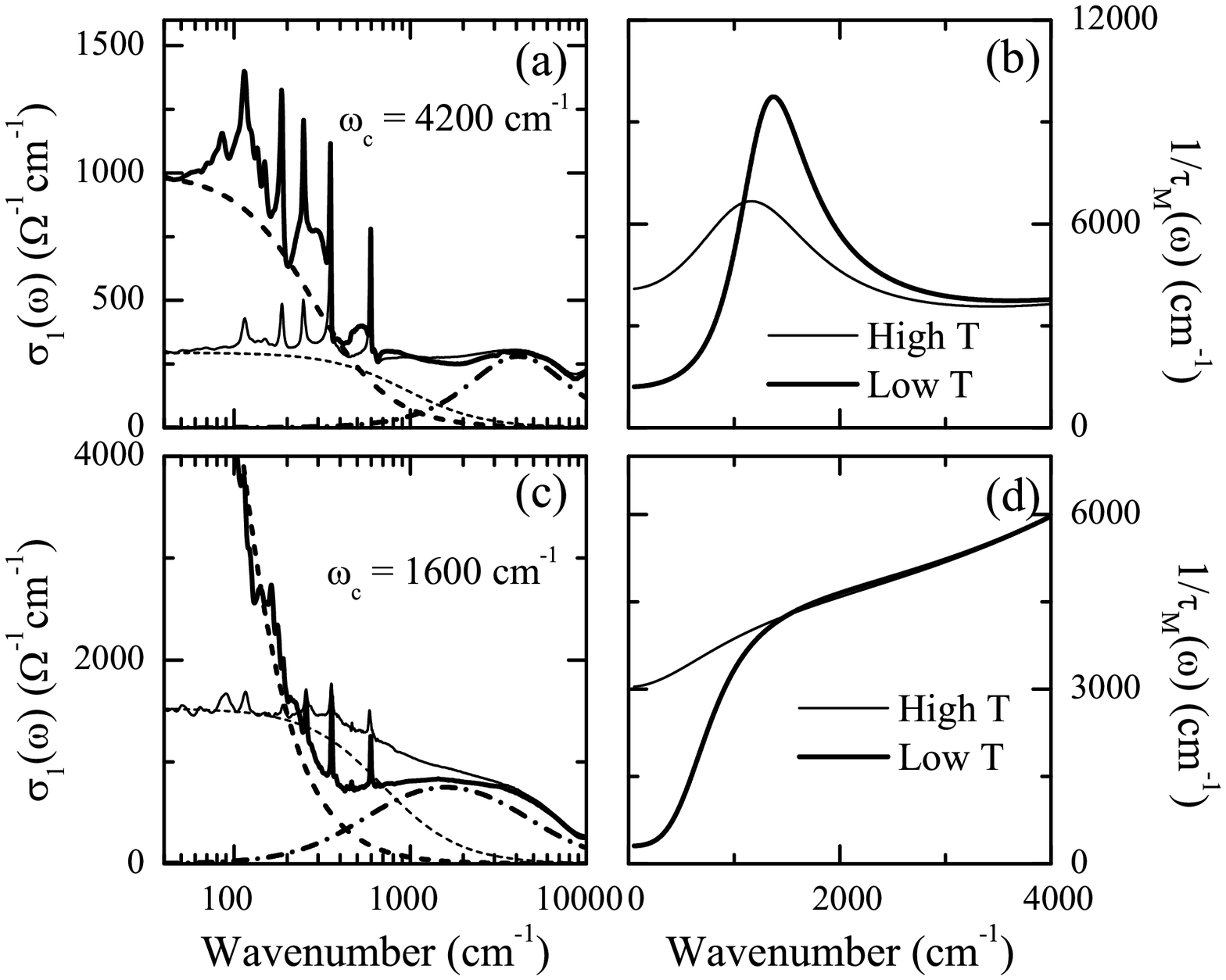}
\caption{Left panels: $\protect\sigma _{1}(\protect\omega )$ for (a) $y=6.35$
and (c) $y=6.65$ crystals. The dotted and the dot-dashed lines represent
contributions due to the Drude mode and Lorentzian oscillator, respectively.
The thick (thin) lines are for low (high) $T$. Fitting parameters are
summarized in Table II. Right panels display modeled scattering rate 1/$%
\protect\tau _{M}(\protect\omega )$ with fitting parameters for (b) $y=6.35$
and (d) $y=6.65$ crystals.}
\end{figure}

A conspicuous aspect of the phase diagram is that the electromagnetic
response of YBCO is radically altered below the crossover temperature $%
T^{\ast }$ (one-component $\rightarrow $ two-component transformation). This
is surprising since the $T^{\ast }$ boundary is rather `soft' given that no
phase transition can be linked to $T^{\ast }$. At the same time, a true
phase transitions leading to the formation of the long range
antiferromagnetism does not trigger significant modifications in the optical
data. This is especially surprising in the context of a sharp onset of
superconductivity near $y$ = 6.40. Also data for both YBCO and LSCO
convincingly show that a certain minimum concentration of dopants is needed
to initiate superconductivity. However, we are unable to recognize
substantial distinctions between the response of AF-ordered and
superconducting crystals apart from mild increase in the oscillator strength
of both coherent and mid-IR contributions to $\sigma _{1}(\omega )$ spectra.

\section{Understanding the optical conductivity trends: single-component
versus multi-component description}

Two principal approaches are commonly used to describe the in-plane
electromagnetic response of high-$T_{c}$ cuprates: \textit{multi-component}
and \textit{one-component} model.\cite{basov04} Within the latter approach
it is assumed that the sole cause of the frequency dependence of $\widetilde{%
\sigma }(\omega )$ is the response of itinerant carriers which acquire
frequency dependent scattering rate $1/\tau (\omega )$ and frequency
dependent mass $m^{\ast }(\omega )$ as the result of strong interactions in
a system. The $1/\tau (\omega )$ and $m^{\ast }(\omega )$ spectra can be
evaluated using the extended Drude model (EDM): 
\begin{equation}
\widetilde{\sigma }(\omega )=\frac{\omega _{p}^{2}}{4\pi }\cdot \frac{1}{%
1/\tau (\omega )-i\omega m^{\ast }(\omega )}\text{,}
\end{equation}%
where $\omega _{p}$ is the plasma frequency usually inferred from the
integration of $\sigma _{1}(\omega )$ up to the energy of the CT excitation.
The multi-component approach (Eq. \ref{eq:drude-lor}) describes the
functional form of the conductivity spectra using a set of (at least two)
Lorentzian oscillators: 
\begin{equation}
4\pi \widetilde{\sigma }(\omega )=\frac{\omega _{p,D}^{2}}{\Gamma
_{D}-i\omega }+\frac{\omega _{p,L}^{2}\omega }{i(\omega _{c}^{2}-\omega
^{2})+\omega \Gamma _{L}}\text{.}  \label{eq:drude-lor}
\end{equation}%
In this equation the first term stands for the Drude response of the free
carriers; $\omega _{p,D}$ is the Drude plasma frequency, and $\Gamma _{D}$
is the scattering rate of the free carriers. The second term stands for the
response of bound charges and has the form of a Lorentzian oscillator
centered at $\omega _{c}$ with a plasma frequency $\omega _{p,L}$ and
scattering rate $\Gamma _{L}$. Particular microscopic scenarios leading to
these terms will be discussed in Section VII.

A debate on both merits and pitfalls of the two scenarios (Eqs. 1 and 2)
goes back to the early days of high-$T_{c}$ superconductivity.\cite%
{book1,quijada99} Here we will focus on the behavior of $1/\tau (\omega
)=\omega _{p}^{2}/4\pi \cdot Re[1/\widetilde{\sigma }(\omega )]$
pertaining to the problem of the pseudogap. The right panels of Fig. 3
uncover the evolution of 1/$\tau (\omega )$ with temperature and doping. The
bottom panels presenting the $1/\tau (\omega )$ data for the $y=6.65$ and $%
=6.75$ crystals are in good agreement with the earlier results for
underdoped YBCO.\cite{basov96} $1/\tau (\omega )$ is nearly linear in $%
\omega $ at room temperature but shows a characteristic threshold structure
near 500 cm$^{-1}$ at $T<T^{\ast }$. It is this depression that is usually
associated with the pseudogap state.\cite%
{basov96,Puchkov96,puchkov96prl,basov02prb} $T^{\ast }$ is as high as 300 K
for the $y$ = 6.65 sample but is reduced down to $\sim $ 200 K for the $%
y=6.75$ compound. In the latter material we also find a parallel offset of
the $1/\tau (\omega )$ for $T>T^{\ast }$. It has been also asserted that
this form of $1/\tau (\omega )$ spectrum is indicative of coupling of QP to
a (bosonic) mode occurring in the vicinity of the threshold structure.\cite%
{carbotte99,Munzar99,Hwang04}

\begin{table}[b]
\caption{Summary of parameters for two-component model spectra $\widetilde{%
\protect\sigma _{M}}(\protect\omega )$ and $1/\protect\tau _{M}(\protect%
\omega )$ spectra shown in Fig. 6. $\protect\omega _{p,D}$ the plasma
frequency and $\Gamma _{D}$ the scattering rate for coherent mode; $\protect%
\omega _{p,L}$ the plasma frequency, $\Gamma _{L}$ the width, and $\protect%
\omega _{c}$ the position of the Lorentzian oscillator; $\protect\omega %
_{p,M}$ is the total plasma frequency for $1/\protect\tau _{M}(\protect%
\omega )$. All parameters are given in cm$^{-1}$. Data in parenthesis are
for the low-$T$ spectra (thin lines in Fig.~6).}%
\begin{ruledtabular}
\begin{tabular}[t]{c|cccccc}
$y$ & $\omega _{p,D}$ & $\Gamma _{D}$ & $\omega _{p,L}$ & $\omega _{c}$ & $%
\Gamma _{L}$ & $\omega _{p,M}$ \\
\hline
6.35 & 4100 & 280(950) & 10850 & 4200 & 7000 & 8500 \\
6.65 & 8000 & 60(700) & 15000 & 1600 & 5000 & 18000%
\end{tabular}
\end{ruledtabular}
\end{table}

New unexpected features of the in-plane electrodynamics are uncovered by the
data for heavily underdoped materials (top right panels in Fig.3). In these
materials the 1/$\tau (\omega )$ spectra at room temperature show a broad
peak around 1000 cm$^{-1}$. With $T$ decreasing, the scattering rate in
far-IR is significantly suppressed, while the peak intensity is enhanced and
the peak position shifts to higher frequencies. The non-monotonic form of
the $1/\tau (\omega )$ observed in all crystals with $y<6.5$ is significant.
Indeed, the negative slope (or peak structure) of the 1/$\tau (\omega )$ at
high frequencies is inconsistent with a single-component description of the
electromagnetic response. This statement become readily apparent, for
example, after a brief inspection of the Allen$^{\prime }$s formula
developed within the framework of the electron-boson scattering:\cite%
{allen71,shulga91} 
\begin{equation}
1/\tau (\omega )=\frac{2\pi }{\omega }\int_{0}^{\omega }d\omega ^{\prime
}(\omega -\omega ^{\prime })\alpha ^{2}F(\omega ^{\prime })+\frac{1}{\tau _{%
\text{imp}}}\text{,}  \label{eq:one}
\end{equation}%
where $\alpha ^{2}F(\omega )$ is the electron-boson spectral function and $%
1/\tau _{\text{imp}}$ is the impurity scattering. Because of the integral
relationship between $1/\tau (\omega )$ and $\alpha ^{2}F(\omega )$ the
slope of the scattering rate spectrum cannot be negative. Therefore, the
non-monotonic dependence of 1/$\tau (\omega )$ at low dopings clearly
indicates a breakdown of the EDM analysis (or single component analysis) to
the heavily underdoped samples. This is hardly surprising given the fact
that for the latter materials the two distinct absorption features are
readily detected in the $\sigma _{1}(\omega )$ spectra. As the doping
increases, the peak structures in the 1/$\tau (\omega )$ spectra are
suppressed and shift to higher frequencies. At doping $\ y\geq 6.5$ the
peaks disappear from $1/\tau (\omega )$ making the slope in the 1/$\tau
(\omega )$ positive at all frequencies. A suppression in the far-IR 1/$\tau
(\omega )$ at lower $T$ is observed at all the dopings and merely reflects
the narrowing of the coherent component in the conductivity data.

We will now demonstrate that the gross features of both $\sigma _{1}(\omega )
$ and 1/$\tau (\omega )$ spectra described above can be qualitatively
reproduced with a two-component scenario for the electromagnetic response.
We modeled the coherent contribution to the conductivity using the Drude
formula with constant scattering rate $\Gamma _{D}$: whereas mid-IR
absorption was accounted for with a single Lorentzian oscillator [Eq. (2)].
Separate contributions due to these two components are displayed with the
dotted and dot-dashed lines in the left panels of Fig. 6. The corresponding
form of the $1/\tau (\omega )$ spectra is presented in the right panels.
First, we fitted the experimental $\sigma _{1}(\omega )$ for $y$ = 6.35 with
the form of Eq. (2). Table II summarizes the fitting parameters. The
parameters of the mid-IR contribution remained unchanged for modeling both
the 293 K and 10 K data; the width of the Drude contribution was reduced to
account for the narrowing of the coherent mode in the low-$T$ spectra. As
clearly seen in Fig. 6(b), the modeled scattering rate $1/\tau _{M}(\omega )$
reasonably reproduces the experimental data [right panel for $y$ = 6.35 in
Fig. 3] including the broad peak structure and its $T$-dependence. We
believe this outcome of the fitting procedure is rather natural since the
two component absorption is evident in the heavily underdoped region
directly in the conductivity data.

\begin{figure}[tbp]
\includegraphics[width=0.4\textwidth]{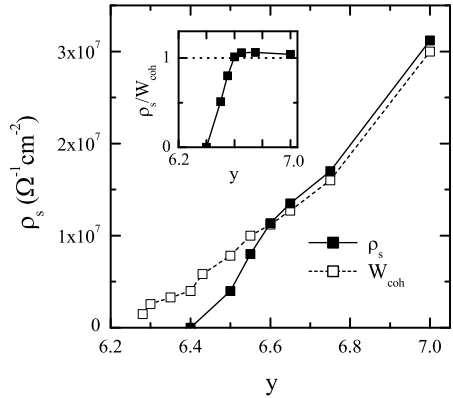}
\caption{Doping dependence of the superfluid density $\protect\rho _{s}$
(filled symbols) and the coherent spectral weight $W_{coh}$ (open symbols)
for $a$-axis data of a series of YBCO single crystals. Inset displays the
ratio of $\protect\rho _{s}$ and $W_{coh}$. }
\end{figure}

Encouraged by the success of the two-component description of the data at
very low dopings we applied Eq. (2) to model the spectra obtained for the $y$
= 6.65 crystal. We found that the frequency dependence of the modeled
spectra 1/$\tau _{M}(\omega )$ as well as their temperature dependence
reproduce the key characteristic features of the experimental data in the
pseudogap state [right panel for $y$ = 6.65 in Fig. 3]. Naturally, the low
frequency suppression of 1/$\tau (\omega )$ is associated with the
development and narrowing of the coherent mode. Only a minute $T$-dependence
of 1/$\tau (\omega )$ at frequencies above 500 cm$^{-1}$ is accounted for by
the mid-IR band which is essentially independent of temperature. We wish to
emphasize nearly linear frequency dependence of the modeled 1/$\tau (\omega
) $ spectra in Fig. 3. This frequency dependence is often regarded as one of
the most anomalous features of the normal state response of of High-$T_{c}$
superconductors. Our modeling shows that this may be a trivial outcome of
the two-component character of the electromagnetic response. Note that in
these calculations we employed the simplest model possible: one Lorentzian
oscillator and a Drude mode with the single temperature dependent parameter $%
\Gamma _{D}$. Obviously, improved fits can be obtained with more flexibility
in the choice of the parameters. Nevertheless, it became possible to
reproduce all important trends in the data using this simplest approach. We
therefore conclude that the two-component model offers a sufficiently
accurate description of the totality of experimental data including the
response in the pseudogap state whereas the single-component approach
clearly breaks down for doping below $y<6.5$.

\section{Nodal metal in the superconducting state}

We now focus on the transformation of the optical conductivity below $T_{c}$%
. A salient feature of all superconducting crystals is a depression of $%
\sigma _{1}(\omega )$ in far-IR region at $T<T_{c}$ with the transfer of the
\textquotedblleft missing\textquotedblright\ spectral weight to
superconducting $\delta $-peak at $\omega =0$. This is in accord with
earlier experimental work reviewed in Ref. 27. Infrared experiment enables
reliable extraction of the superfluid density $\rho _{s}$ from the optical
constants.\cite{Dordevic02} In Fig.7 we plot the doping dependence of the
superfluid density obtained for the series of samples that we have
investigated. One find that $\rho _{s}$ does not exceed the coherent
contribution to the conductivity $W_{coh}$. The latter weight can be
evaluated from integration of the conductivity up to 600 cm$^{-1}$: the
frequency range where the coherent component dominates in the optical
response.

One interesting observation pertains to the frequency dependence of the $%
\sigma _{1}(\omega )$ for $y$ = 6.65 and 6.75 obtained just above $T_{c}$
and at 10 K. The form of these spectra is very similar and the only
difference is in the diminished spectral weight in the 10 K data fully
accounted for by the area under the $\delta $-function.\cite{Basov01,Homes04}
New results reported here for even lower dopings: $y$ = 6.50 and 6.55 (Fig.
8) uncover several unexpected features. We present these spectra on the log
scale in order to clearly display both the coherent contribution to the
conductivity as well as mid-IR band. Notably, the two absorption structures
are well separated one from each other at these lower doping. This
circumstance allows one to evaluate the role of both of these conductivity
channels in the formation of superconducting condensate through the
examination of spectral features rather than through the evaluation of the
integrated spectral weight as in Fig. 7. The superconducting state spectra
show that the spectral weight associated with the coherent component has
been significantly diminished. Changes of the optical conductivity are
primarily confined to the frequency range determined by the magnitude of the
scattering rate in the $T\sim T_{c}$ curves. We do not observe any
significant depletion of the conductivity associated with the mid-IR band at 
$T<T_{c}$. We therefore conclude that the dominant contribution to the
superconducting condensate originates from the coherent quasiparticles on
the Fermi arc.

\begin{figure}[tbp]
\includegraphics[width=0.35\textwidth]{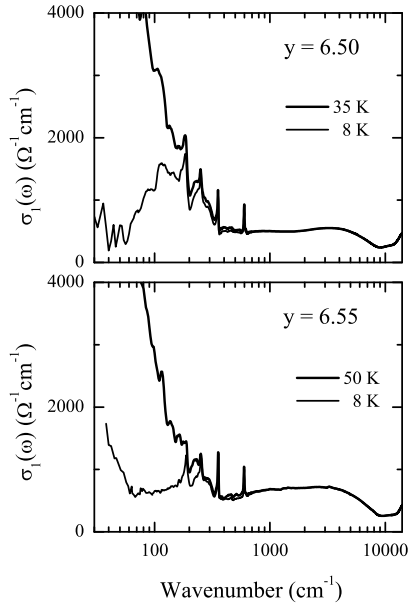}
\caption{Spectra of the $a$-axis conductivity $\protect\sigma _{1}(\protect%
\omega )$ for $y$ = 6.50 [top panel] and 6.55 [bottom panel] at $T\ll T_{c}$
(thin lines) and $T\sim T_{c}$ (thick lines). }
\end{figure}

In the superconducting state spectra plotted in Fig. 8 one can identify a
gap-like threshold structure in $\sigma _{1}(\omega )$ at $\sim $ 100 cm$%
^{-1}$ for $y$ =6.50 crystal and at $\sim $ 220 cm$^{-1}$ for $y$ = 6.55
sample. These features are not seen at $T\geq T_{c}$ and gradually develop
with temperature lowering below the critical value. We assign this structure
with superconducting energy gap. This assignment is warranted since these
low-energy structures are well separated from all other absorption features.
We also point out that the superconducting state conductivity remains finite
down to the lowest energies. This is in the qualitative agreement with the
expected theoretical behavior for dirty $d$-wave superconductor.\cite%
{carbotte95} Alternatively, finite residual conductivity below the
superconducting gap can be attributed to the tail of mid-IR band extending
down to the lowest frequencies and/or inhomogeneous superconducting
condensate\cite{orenstein03}.

\section{Magnetic resonance in the in-plane charge dynamics of YBCO}

An important aspect of charge dynamics of cuprates is the possibility of QP$%
^{\prime }$s coupling to collective modes. A well-known example is the
so-called Holstein bands arising in systems with strong electron-phonon
interaction.\cite{allen71,holstein54,joyce70,timusk76} It was suggested
early on that deviations of the in-plane optical conductivity of cuprates
from conventional Drude form may originate from coupling to a bosonic mode.%
\cite{thomas88} Later analysis of the $\widetilde{\sigma }(\omega )$\cite%
{carbotte99,Munzar99,normal98} has indicated that the relevant mode in
cuprates may be related to the so-called 41 meV magnetic resonance\cite%
{dai01,fong00} observed in neutron scattering experiments. One appeal of the
strong coupling approach is in its well-defined predictions, not only for IR
experiments,\cite{carbotte99,Hwang04} but also for data generated with other
spectroscopic techniques including angle resolved photoemission spectroscopy
(ARPES)\cite{johnson01,abanov99} as well as tunneling.\cite%
{zasadzinski01,abanov00} At least in the case of the optimally doped YBCO
and Bi$_{2}$Sr$_{2}$CaCu$_{2}$O$_{8+\delta }$ (Bi2212) superconductors this
analysis indeed provides a consistent account of the above mentioned
spectroscopies based on the idea of QP coupling to a magnetic resonance.
Recently, the role of magnetic resonance in QPs dynamics has been
challenged. An examination of ARPES experiments has suggested that the
totality of data is better described in terms coupling to phonons\cite%
{lanzara01} and not to magnetic excitations. This claim is not supported by
the IR studies of isotopically substituted YBCO which show no isotope effect
for the feature in question.\cite{wang02,bernhard04}

\begin{figure}[tbp]
\includegraphics[width=0.48\textwidth]{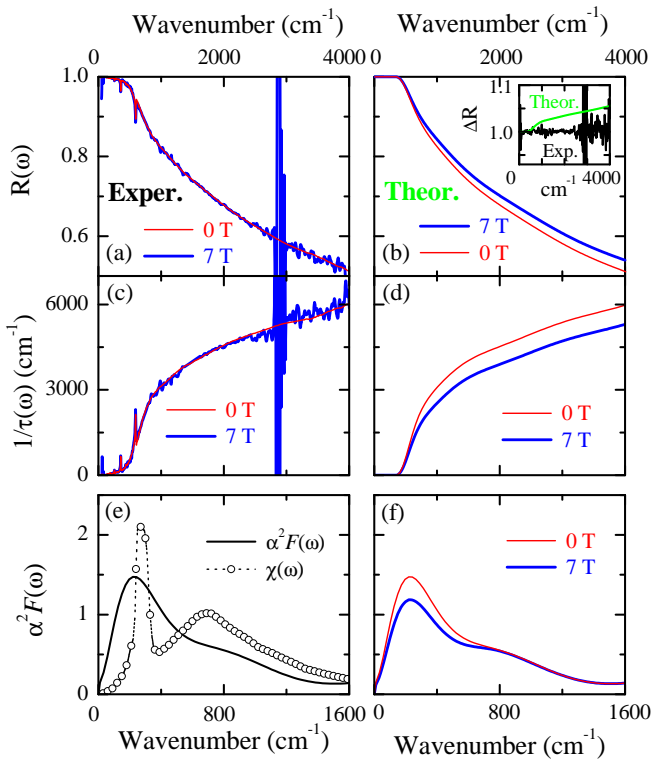}
\caption{(color online) Low temperature reflectance spectra (top) $R$($%
\protect\omega $), (middle) 1/$\protect\tau (\protect\omega )$ spectra and
(bottom) $\protect\alpha ^{2}F(\protect\omega )$ data for $y$ = 6.65 YBCO
single crystal at all 5 K. Red (thin) lines: \textbf{H} = 0 T; blue (thick)
lines: \textbf{H }= 7 T. Left panels: experimental results. Right panels:
model spectra calculated using the protocol described in the text. The $%
\protect\alpha ^{2}F(\protect\omega )$ data extracted from the H = 0 T and 7
T spectra are nearly identical. Inset in (b): $\Delta R(\protect\omega ,$%
\textbf{H}$)=R(\protect\omega ,7$ T)/$R$($\protect\omega $, 0 T). Sharp
spikes in the high field spectra are due to absorption in the windows of our
cryostat. To calculate $\protect\alpha ^{2}F(\protect\omega )$ we used $%
\Delta =180$ cm$^{-1}$. Also shown with open symbols in panel (e) is the
spin susceptibility $\protect\chi (\protect\omega )$ from the INS data
reported in Ref. \protect\cite{dai99} for $y$ = 6.6 ($T_{c}$ = 62.7 K)
single crystal. The $\protect\chi (\protect\omega )$ spectrum is similar to
the experimental result for $\protect\alpha ^{2}F(\protect\omega )$ obtained
from the inversion of IR data.}
\end{figure}

Insights into strong coupling effects may be gained from studies of the QP
dynamics in magnetic field. The rationale for this approach is provided by
the work of Dai \textit{et al}. who discovered that the intensity of the
magnetic resonance in the $y$ = 6.6 YBCO crystal ($T_{c}$ = 62.7 K) is
suppressed by 20 $\%$ in 6.8 T field applied along the $c$-axis \cite{dai00}%
. Other candidate excitations including phonons, or the continuum of spin
fluctuations, are unlikely to be influenced by a magnetic field of similar
modest magnitude. For this reason an exploration of the field-induced
modifications of the electronic self-energy enables a direct experimental
inquiry into the role of the magnetic resonance in QP properties and on a
more general level, into an intricate interplay between superconductivity
and magnetism in cuprates.

In order to quantify the magnitude of possible \textbf{H}-induced changes in
the reflectivity spectra we adopted the following procedure. We first
extracted the spectral function $\alpha ^{2}F(\omega )$ from the zero field
data for $y$ = 6.65. The detailed method is published elsewhere.\cite%
{dordevic04} An extracted spectrum shows fair agreement with experimental
results for the spin susceptibility $\chi (\omega )$ obtained from inelastic
neutron scattering (INS) experiments [open symbols in Fig. 9(e)].\cite{dai99}
Specifically, both the peak at 250 cm$^{-1}$ and a broad background of the
INS data is reproduced through this analysis. We then reduced the intensity $%
\alpha ^{2}F(\omega )$ by 20 \% without modifying the broad background in
accord with INS measurements.\cite{dai00} Using the spectral function with
the suppressed intensity we calculated $1/\tau (\omega ,$ 7 T$)$,\cite%
{dordevic04} and also $m^{\ast }(\omega $, 7 T$)$ with the help of
Kramers-Kronig analysis. Finally, a combination of $1/\tau (\omega ,$ 7 T$)$
and $m^{\ast }(\omega $, 7 T$)$ allowed us to generate the reflectance
spectrum $R(\omega ,$ 7 T$)$ [blue (thick) line in Fig. 9(b)]. Comparing
this final output of modeling with the experimental curve for \textbf{H} = 0
one finds that the effect of the applied magnetic field is rather small in
the far-IR but exceeds 5 \% at frequencies above 800 cm$^{-1}$. Moreover,
anticipated changes of reflectance exceed the conservative estimate for the
uncertainty of $R(\omega ,$\textbf{H}) in a new apparatus we have developed
for infrared magneto-optics\cite{magnetooptic} and therefore could be
readily detected.

Representative results are displayed in Fig. 9. Here we plot the raw
reflectance spectra measured at $T$ = 5 K for $y=6.65$ crystals. The spectra
for the latter material are in good agreement with the earlier studies of
YBCO with similar oxygen content.\cite{basov96}. In our magneto-optics
apparatus we are capable of measuring the absolute values of reflectivity in
the magnetic field.\cite{magnetooptic} For $y$ = 6.65 system we found that
the field-induced changes of the reflectivity are negligibly small either
under zero-field cooling or under in-field cooling conditions. The same
instrument was successfully used to monitor transitions between the Landau
levels in graphite which produce only weak changes of $R(\omega )$
comparable to anticipated effect in YBCO.\cite{magnetooptic} We also
repeated measurements for $y$ = 6.50 crystal and did not detect significant
field-induced changes of reflectance in mid-IR. These results call for a
revision of the prominent role of magnetic excitations in QPs dynamics.

\section{Discussion}

\subsection{Two-component quasiparticles dynamics of the nodal metal and the
pseudogap crossover}

The key experimental finding of this work is that the gross features of the
low-temperature electrodynamics in YBCO are adequately described within the
two-component model (Eq.~2). This simple description holds throughout an
extended region of the phase diagram from AF-ordered phases to $d$-wave
superconductor. The separation between the coherent Drude contribution to
the conductivity and mid-IR band is most evident in the low-$T$ response at
very low dopings. This aspect of electrodynamics is not specific to YBCO but
is also common to the LSCO system\cite{Padilla04,uchida91} as well as for
the electron-doped Nd$_{2-x}$Ce$_{x}$CuO$_{4}$ system.\cite%
{onose04,zimmers04,wang04} As doping increases the spectral weight
associated with both the coherent component and mid-IR band is enhanced
(Fig. 10). Moreover, the mid-IR band systematically softens with increasing
doping. The net result of these effects is that the two contributions merge
and can no longer be separated beyond a certain doping level ($y>$ 6.75 in
YBCO and $x>$ 0.125 in LSCO). We therefore conclude that the doping
dependent evolution of optical spectra appears to reflect a generic property
of cuprates.

Our experiments unequivocally show that at the pseudogap boundary of the
phase diagram the charge dynamics of the CuO$_{2}$ plane experiences a
crossover from a single-component type to two-component response of the
nodal metal. Moreover, we find that at $T<T^{\ast }$ both transport and
spectroscopic properties are consistent with the Fermi liquid (FL) theory.%
\cite{pines66} This conjecture is supported by the Drude frequency
dependence of the optical conductivity as well as by the $T^{2}$ form of the
resistivity\cite{ando04} plotted in Fig. 11. A prerequisite for the FL
theory is well-defined quasiparticle excitations. The existence of such
quasiparticles at $T$ $<T^{\ast }$ is in accord with relatively large values
of the electronic mean free path extracted from the analysis of the coherent
component in the conductivity.\cite{Yslee04} Further evidence for
well-defined quasiparticles at $T<T^{\ast }$ is provided by observations of
bi-layer splitting effects both by ARPES\cite{borisenko04} and by IR studies
of the $c$-axis response.\cite{dodervic04-1} Thus the totality of the
experimental data suggests the Fermi liquid nature of the nodal metal.

Attributes of the FL dynamics of the nodal metal are most vivid in the
temperature-doping parameter space where the coherent contribution is
energetically separated from the incoherent mid-IR band. Photoemission
experiments for the LSCO system conclusively show that this regime is
realized when most of the Fermi surface is gapped and the only remaining
portion is the arc formed around the nodal points.\cite%
{Yoshida03,Ino02,damascelli03} However, the FL hallmarks can no longer be
identified when the two-component behavior is terminated at the pseudogap
boundary and the large Fermi surface is recovered. A corresponding feature
of the optical data is the merger between Drude and mid-IR contributions,
which is adequately described with the anomalous $\omega $-dependence in 1/$%
\tau (\omega )$, referred to as non-Fermi liquid.\cite{book1,marel03} Hall
measurements for LSCO show a dramatic enhancement of the effective number of
carriers participating in transport at the same boundary. In the earlier
publication we have established a quantitative consistency between IR and
Hall data.\cite{Padilla04,ando04} Unfortunately, ARPES and high-temperature
resistivity/Hall data are available only for LSCO system. However, close
similarity between IR results for both LSCO and YBCO (Fig. 10) prompt us to
conclude that the above trends may be reflecting intrinsic properties of
weakly and moderately doped CuO$_{2}$ planes that are valid irrespective of
a particular host material.

\begin{figure}[tbp]
\includegraphics[width=0.48\textwidth]{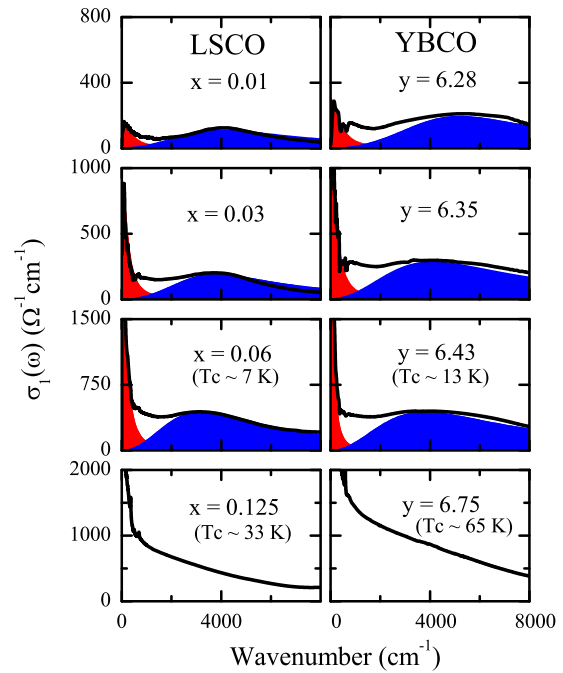}
\caption{(color online) Evolution of the optical conductivity with doping in
LSCO (left panels) and YBCO (right panels) series from Ref.\protect\cite%
{Padilla04}. Data at 10 K is presented for non-superconducting crystals;
data at $T\simeq T_{c}$ are shown for superconducting materials (at $T\simeq
T^{\ast }$ for $y$ = 6.75). For clarity, phonons were removed from all
spectra by fitting them with the Lorentzian oscillators. The response of
weakly doped samples shows a Drude like behavior at low frequencies followed
by a resonance in mid-IR. Red (light gray) and blue (dark gray) areas
represent the Drude and mid-IR absorption modes in Eq.~(2), respectively.}
\end{figure}

\begin{figure}[tbp]
\includegraphics[width=0.37\textwidth]{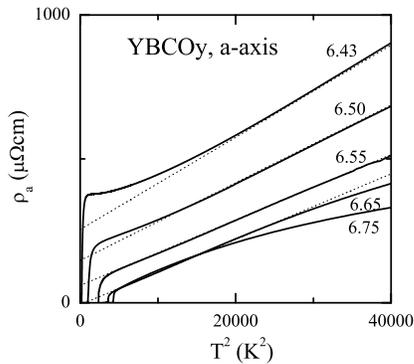}
\caption{$T^{2}$-dependence in dc resistivity $\protect\rho _{\text{dc}}(T)$
in the YBCO series. In pseudogap state ($y$ = 6.65 and 6.75) the $T^{2}$%
-dependence is shown clearly. Even in heavily underdoped region $\protect%
\rho _{\text{dc}}(T)$ exhibits $T^{2}$-dependence at moderately high
temperatures. It is noted that the cotangent of the Hall angle, cot$\Theta
_{H}$, also shows the $T^{2}$ dependence in the corresponding region, which
implies that the scattering rate should follow the $T^{2}$ dependence. For
detailed discussion, see Ando \textit{et al}., Phys. Rev. Lett. \textbf{92},
197001 (2004).}
\end{figure}

\subsection{The origin of mid-IR band}

The discussion in the previous subsection shows that the mid-IR band
universally found in all high-$T_{c}$ superconductors is intimately involved
in the pseudogap phenomenology and specifically in transport properties at
high temperatures.\cite{Padilla04,ando04} Moreover, the softening of the
mid-IR band with doping resembles the decrease of pseudogap temperature $%
T^{\ast }$ as shown in the bottom panel of Fig. 5. In view of these
preeminent roles of the MIR resonances it is prudent to inquire into the
physics underlying this absorption feature. The form of the 1/$\tau (\omega
) $ spectra for weakly doped YBCO resembles that of systems with spin-
and/or charge density wave.\cite{basov02prb,lee01} This is a likely
possibility given overwhelming evidence for spin/charge self-organization
effects in weakly doped phases.\cite{carlson02} Recently, charge ordering
patterns have been directly detected using scanning tunneling spectroscopy.%
\cite{davis04,yazdani04} If the position of the mid-IR is chosen to
characterize the magnitude of the electronic gap we find that $\omega _{%
\text{mid}}=7-9$ $k_{B}T^{\ast }$. This ratio is quite common for charge
density wave states in solids.\cite{gruner}

Several other possible scenarios for mid-IR band are worthy of our
attention. A detailed analysis of the electronic structure of doped Mott
insulators predicts several forms of bound states within the Mott-Hubbard
and/or charge transfer gaps.\cite{dagotto94} Then interband transitions
involving these states may give rise to the observed effects in mid-IR.
Recent ARPES results have reported that apart from the Fermi arc in the
nodal region, the so-called flat band is formed around ($\pi $, 0$)$ of
Brillouin zone at $\sim $ - 0.2 eV in heavily underdoped region and rises up
to the Fermi level with the increased doping.\cite{Yoshida03,Ino02}
Interestingly, this doping dependence of the flat band is reminiscent of the
softening of the mid-IR absorption. A different view on the nature of the
mid-IR structure is given by Lorenzana and Sawatzky who argued that this
feature is due to a quasibound state of two magnons coupled to an optical
phonon.\cite{lorenzana95} Alternatively, the multi-component response
including the mid-IR absorption might be attributed to the real space
electronic inhomogeneity occurring due to embedding of metallic regions in
an insulating host. Near the percolation threshold the optical conductivity
of such a system shows a Drude response at low energies followed by a
featureless background.\cite{emery93,emery95,kivelson04} Yet another
possibility is that the mid-IR absorption is produced by an incoherent band
formed by the strong interaction of carriers with phonons\cite{lupi98} or
spin fluctuation.\cite{Salk02} Finally, Leggett\cite{leggett99} and Turlakov%
\cite{turlakov03} discussed the plasmonic nature of the mid-IR spectral
feature. Within this latter scenario sizable changes in mid-IR spectra below 
$T_{c}$ were predicted.\cite{leggett99} Experimentally, the temperature
dependence of the mid-IR absorption is rather weak.\cite{ellipsometry} While
further studies on the origin of the mid-IR absorption are needed, our
findings strongly suggest that any explanation should take into account the
correlation between the formation of the mid-IR absorption and the
development of the pseudogap.

\subsection{Self-energy effects in the quasiparticles dynamics.}

As pointed out in Section VI, reflectivity measurements performed in \textbf{%
H }$\parallel c$ field do not show noticeable changes at the mid-IR
frequencies calling for a critical re-examination of the roles of magnetic
excitations in quasiparticle dynamics. Specifically, IR data reported in
Fig.~9 challenge the relevance of the resonance detected in the INS
experiments to the electronic self-energy effects in mid-IR. An issue of
whether or not the magnetic mode seen by the INS is capable to seriously
impact the electronic self-energy in view of only small intensity of the
resonance has been contested in the literature.\cite{kee02,abanov02,eremin04}
Our new results reported in Fig.~9 in conjunction with the INS experiments
in high magnetic field challenge a preeminent role of magnetic resonance in
electrodynamics of cuprates. More importantly, the self-energy
interpretation of the IR data or the single-component approach appears to be
in conflict with the main experimental finding of this work: the
two-component nature of the electromagnetic response. It is therefore
imperative to take into account this other contribution before inquiring
into the role of self-energy in the coherent component of the conductivity.

As a note of caution it must be stressed that marked effects are not
necessarily expected if the magnetic resonance broadens in the applied
field. Since it is the integrated weight of the magnetic mode that is
relevant for self-energy a simple smearing of the resonance is unlikely to
significantly modify IR data. In this context we stress that INS results of
Dai \textit{et al}. suggest a reduction of the intensity of the mode and not
its broadening.\cite{dai00} If this latter behavior is confirmed by future
neutron studies, our observed behavior is incompatible with the spin exciton
interpretation of the INS resonance mode.

\section{Summary and outlook}

We investigated the electromagnetic responses of a prototypical high-$T_{c}$
cuprate YBCO in a broad region of the phase diagram from the AF to the
pseudogap state ($y$ = 6.28 - 6.75). We focused on the analysis of the nodal
metal phase that is characterized by a clear energy separation between the
low-energy electronic states responsible for Drude conductivity and higher
energy excitations producing mid-IR structure in the optical data. We
emphasized close parallels in the IR data for YBCO and LSCO systems and
concluded that the two-component characteristics may be generic for high-$%
T_{c}$ cuprates. At least in the case of LSCO, the two-component nature of
the electronic excitations is also consistent with both the ARPES and
transport results.

A combination of transport and IR experiments has allowed us to identify
several hallmarks of the Fermi liquid in the properties of the nodal metal.
Interestingly, this rather conventional electronic behavior characterized
with the high electronic mobility\cite{Ando01} and relatively low effective
mass\cite{Padilla04} extends to AF ordered phases. These latter findings
clearly show that the transition from a nodal metal to a Mott insulator is
of \textquotedblleft vanishing carrier number\textquotedblright .\cite%
{imada98} The above experiments also suggest that transport in nodal metals
may be governed by excitations topologically compatible with an
antiferromagnetic background. Many of the doping trends reported here are
consistent with the projected wave functions approach.\cite{paramekanti01}
One conjecture reconciling anomalous trends seen in weakly doped cuprates is
that the local environment of mobile charges in these systems remains
unaltered with doping and it is only the phase space occupied by hole rich
regions that is progressively increasing.

We show that the pseudogap state in the generic phase diagram is associated
with a crossover from the two-component conductivity of a nodal metal to the
single component response at $T>T^{\ast }$. In the vicinity of the pseudogap
boundary the effective number of charge carriers contributing to transport
and optics is enhanced.\cite{Padilla04,ando04} This indicates that the
functional form of the dc resistivity of cuprates $\rho _{\text{dc}}(T)$ is
governed not only by the relaxation processes but also by temperature
dependent number of carriers. Even though the mid-IR band is contributing to
transport at high temperature, the low-temperature properties of cuprates
are dominated by nodal quasiparticles. In superconducting crystals the vast
majority of the superfluid spectral weight is produced by condensation of
nodal quasiparticles.

High-field magneto-optics experiments pose challenges for the interpretation
of the IR spectra in terms of fermionic self-energy effects prompted by
coupling of quasiparticles to a neutron resonance. In view of the
two-component response documented here it is imperative to take into account
the mid-IR contribution before the low-frequency data is employed to search
for strong coupling effects. However, this task is connected with ambiguous
procedures to remove the mid-IR contribution from the data. We therefore
have not attempted this analysis here. We note that the success of a simple
Drude+Lorentzian description of the data (Fig. 6) indicates that the
self-energy effects in optics may be rather weak. It is therefore of
interest to re-examine the self-energy correction in the results obtained
with other spectroscopies. Measurements in magnetic field are of high
interest in this context. While it may be impossible to carry out such
experiments in the case of photoemission studies, tunneling measurements
appear to be well suited for this task.

\begin{acknowledgments}
We acknowledge G. Blumberg, K.S. Burch, J.P. Carbotte, A. Chubukov, T.
Timusk, and J.M. Tranquada for helpful discussion. This research was
supported by the US department of Energy Grant and NSF.
\end{acknowledgments}

\end{document}